# Critical Systems Development Using Modeling Languages (CSDUML'04): Current Developments and Future Challenges (Report on the Third International Workshop)

Jan Jürjens[1,*], Eduardo B. Fernandez[2],
Robert B. France[3], Bernhard Rumpe[4], and Constance Heitmeyer[5]

[1] Software & Systems Engineering, Dep. of Informatics, TU Munich, Germany
[2] Dep. of Computer Science and Engineering, Florida Atlantic University, USA
[3] Computer Science Department, Colorado State University, USA
[4] Software Systems Engineering, TU Braunschweig, Germany
[5] Naval Research Laboratory, USA

**Abstract.** We give a short report on the contributions to and some discussions made and conclusions drawn at the Third International Workshop on Critical Systems Development Using Modeling Languages (CSDUML'04).

## 1 Introduction

A *critical system* is a system in which compelling evidence is required that the system satisfies critical properties, such as real-time, safety, security, and fault-tolerance properties. The construction of high-quality critical systems, e.g., avionics systems, life-critical medical systems, weapons systems, and control systems for nuclear power plants, can be enormously difficult and costly. In recent years, many critical systems have been developed, and deployed which do not satisfy critical requirements. This has led in many cases to catastrophic system failures.

Part of the difficulty of critical systems development is that producing compelling evidence of the system's correctness can be enormously expensive. Due to high costs, producing detailed system specifications and designs along with evidence that these artifacts satisfy critical properties is normally avoided. Using UML to construct a system model that satisfies critical properties may help lower these costs since UML models are easy to understand, are amenable to mechanized analysis to check critical properties, and can be used to synthesize correct, executable code.

The workshop series on "Critical Systems Development Using Modeling Languages (CSDUML)" aims to gather practitioners and researchers to contribute to overcoming the challenges one faces when trying to exploit these opportunities. The previous editions of the series were the CSDUML'02 satellite workshop of the UML'02 conference in Dresden (Germany) and the CSDUML'03 satellite workshop of the UML'03 conference in San Francisco. Both had been very successful satellite workshops of the UML conferences. The workshop report at hand now gives an overview on the contributions for and outcomes of the CSDUML'04 workshop, which took place on October 12, 2004, as part of the UML'04 conference (October 10 – 15, 2004, in Lisbon, Portugal). It was again organized in cooperation with the pUML (precise UML) group and the working group on Formal Methods and Software Engineering for Safety and Security (FoMSESS) of the German Computer Society (GI).

---

* http://www4.in.tum.de/~juerjens . Supported within the Verisoft Project of the German Ministry for Education and Research.

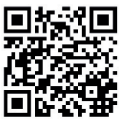

In the following, we first give an overview on the various contributions to the workshop. We will then attempt to draw some conclusions on the current state of the art and future challenges in the area of the workshop.

## 2  Contributions

The workshop featured an invited talk with the title "On the Role of Tools in Specifying the Requirements of Critical Systems" by Constance Heitmeyer (head of the Software Engineering Section of the Naval Research Laboratory's Center for High Assurance Computer Systems and one of the internationally leading experts in the formal specification and formal analysis of software and system requirements and of high assurance software systems). Furthermore, there was a panel with the title "Providing tool-support for critical systems development with UML: Problems and Challenges" consisting of distinguished experts, which created some lively discussions on the subject.

For contributed presentations, out of a number of high quality papers submitted to the workshop, seven were selected to be presented in talks at the workshop and included as full papers in the proceedings. Three additional papers were selected to be presented as short talks and included as short papers. Furthermore, there were six posters presented at the workshop, which are included in the proceedings as abstracts. The highly selective acceptance rate again kept the workshop focused, and at a high level of quality, while allowing sufficient time for discussion.

### C. Heitmeyer (Center for High Assurance Computer Systems, Naval Research Laboratory): On the Role of Tools in Specifying the Requirements of Critical Systems (Invited Talk)

In 1978, a group of researchers led by Dave Parnas developed a tabular notation for specifying software requirements called SCR (Software Cost Reduction) and used the notation to specify the requirements of a mission-critical program, the Operational Flight Program for the A-7 aircraft. Since then, the requirements of many critical programs, including control software for nuclear power plants and other flight programs, have been specified in SCR. To support formal representa-tion and analysis of software requirements, NRL has developed a state machine model to define the SCR semantics and built a suite of tools based on this semantics for checking requirements specifications for properties of interest. Such tools are especially valuable for specifying and analyzing the requirements of software systems where compelling evidence is required that the system satisfies critical properties, such as safety and security properties. This talk described the many different roles that formally based software tools can play in debugging, verifying, and validating the requirements of critical software systems. The author's recent experience and lessons learned in specifying the requirements of a security-critical cryptographic system and two software components of NASA's International Space System were also described."

**R. B. France (Colorado State University), C. Heitmeyer (CHACS, NRL), H. Hußmann (LMU Munich), F. Parisi-Presicce (George Mason University and Univ. di Roma La Sapienza), A. Pataricza (Budapest University of Technology and Economics), and B. Rumpe (TU Braunschweig): Providing Tool-Support for Critical Systems Development with UML: Problems and Challenges (Panel)**

Among the panelists, and also the other workshop participants, a number of controversial issues were discussed. A selection of the discussion points and some of the opinions are given below.

**Sufficient Precision of UML for Cricitical Systems Development.** A range of different opinions were expressed concerning the question whether the UML is presently, or ever will be, sufficiently precisely defined to be suitable for critical systems development. The issue was raised by panelist Heinrich Hußmann who expressed doubt that this will ever be the case. Other workshop participants suggested that with UML as a family of languages, at least a core of UML might be defined in a precise way. Others expressed the concern that precision is not only necessary for critical systems development, but for any kind of advanced tool-supported use of UML, since any aspects of UML that should be supported by tools firstly have to be defined precisely enough to put one in the position to provide the tool-support.

**Role of Tools for Researchers.** Panelist Bernhard Rumpe raised the discussion point that researchers working on tool support for UML should consider the question what the intended use of the tools developed should be. This could start from tools as a means to validate one's ideas with respect to correctness, implementability, feasibility, scalability etc.. It could include the use of tools to demonstrate to academic peers the usefulness of one's concepts or tool-buidling as an activity helping to focus the activities of one or more research groups. A more ambitious goal would be the use as an "in-house" tool by the developers in their own projects. Finally, one could try to actually sell the tools to others who would like to use them (or give them away for free but at least sell support services). Again, a range of opinions were expressed. Bernhard Rumpe expressed concern that with academic tools intended to be used by others, this might incur a considerable effort in support and maintenance in the long run. On the other hand, panelist Connie Heitmeyer suggested that it is an important goal to try to put tools to use, for example to achieve technology transfer from research to industrial practice.

**Other Issues.** As part of the general discussion, several other points were raised. For example, panelist Connie Heitmeyer proposed for discussion whether UML is ideally suited specifically for describing critical requirements (as opposed to the design of critical systems), or whether a different notation might be required for that.

### P. Conmy and R. Paige (University of York): Using UML OCL and MDA to Support Development of Modular Avionics Systems

A Model Driven Architecture approach to the development of Integrated Modular Avionics Systems is explained. For this, the required system behavior is captured in a Platform Independent Model using UML models that are extended with OCL constraints. The model and the constraints are then transformed to a Platform Specific Model. The paper discusses potential benefits and difficulties.

### M. Huhn, M. Mutz, and B. Florentz (TU Braunschweig): A Lightweight Approach to Critical Embedded Systems Design Using UML

The paper presents a pragmatic approach to the UML based design of critical systems that has been applied in the automotive domain. For that, it focuses on so-called lightweight formal methods auch as automated static analysis and validation of dynamic behaviour by simulation (although there is a potential for also incorporating a fully formalized model analysis). The paper also presents a tool with binding to commercial CASE tools used in industry.

### J. Tenzer (University of Edinburgh): Exploration Games for Safety-Critical System Design with UML 2.0

The paper presents an approach which aims to provide a smooth transition from informal UML design models to the kind of precise specifications that are needed in the formal verification of critical systems. The approach is based on exploration games played by the modeler to detect flaws and determine sources of unacceptable imprecision. As part of the game, the design is then improved. The paper discusses these ideas at the hand of UML 2.0 activity diagrams and state machines using a small critical system and gives an outlook on planned tool-support.

**R. Heldal (Chalmers University of Technology), S. Schlager (University of Karlsruhe), and J. Bende (Chalmers University of Technology): Supporting Confidentiality in UML: A Profile for the Decentralized Label Model**

This work has as its goal to incorporate a decentralized label model into the UML by defining a UML profile. The profile allows one to specify the confidentiality of data in UML models by annotating classes, attributes, operations, values of objects, and parameters of operations. From the annotated, code in the Java extension Jif (Java information flow) can be generated in a way which guarantees that the confidentiality constraints are not violated.

**S. H. Houmb (Norwegian University of Science and Technology), G. Georg, R. B. France, and D. Matheson (Colorado State University): Using Aspects to Manage Security Risks in Risk-Driven Development**

The approach presented in this paper extends the CORAS framework, which is an integrated risk management and system development process for security-critical systems based on AS/NZS 4360, RUP, and RM-ODP. In particular, it now makes use of aspects to specify security risk treatment options and to implement security mechanisms. One thus gets an aspect-oriented risk-driven development process where security requirements can be identified in each development phase. The requirements can be treated by making use of the aspects, which facilitates development and evaluation of security treatment options as well as system evolution.

**M. V. Cengarle (TU Munich) and A. Knapp (LMU Munich): UML 2.0 Interactions: Semantics and Refinement**

This paper is concerned with High-Level Message Sequence Charts (HMSCs) which in version 2.0 are newly integrated into UML for interaction modelling. More specifically, it considers the possibility of writing negated specifications that was introduced at the opportunity, which can be used to rule out behaviour from implementations. The paper puts forward a trace-based semantics for UML 2.0 interactions that captures the standard composition operators for UML 2.0 interactions from HMSCs, and also the added negation and assertion operators. Based on that, several alternatives for treating negation in interactions can be discussed. The proposed semantics determines whether a trace is positive, negative, or inconclusive for a given interaction. The paper also defines notions of implementation and refinement for sets of interactions based on this.

**T. Massoni, R. Gheyi, and P. Borba (Federal University of Pernambuco): A UML Class Diagram Analyzer**

The presented work deals with the automated analysis of UML models which include OCL constraints. More specifically, it presents an approach for the au-tomated analysis of UML class diagrams, based on the formal object-oriented modeling language Alloy. It allows us to use Alloy's tool support for class diagrams, by applying constraint solving to automatically find valid snapshots of models. The aim of this automation is in particular to support the identification of inconsistencies or under-specified models of critical software.

**I. Johnson (VT Engine Controls); C. Snook, A. Edmunds, and M. Butler (University of Southampton): Rigorous Development of Reusable, Domain-Specific Components, for Complex Applications**

This paper uses a UML profile called UML-B to develop failure management systems in a model-based refinement style. It thus aims to provide rigorous validation and verification in the presence of systems in evolution. The UML-B profile can be translated into a formal notation called B-method using the U2B translation tool. It includes a constraint and action language to assist behavioral modeling. The aim is thus the reuse of reliable, domain-specific software components, in particular in avionics for safety-critical airborne systems.

**Z. Dwaikat (George Mason University and Cigital) and F. Parisi-Presicce (George Mason University and Univ. di Roma La Sapienza): From Misuse Cases to Collaboration Diagrams in UML**

The research presented here aims to integrate the concepts of misuse and abuse cases from software security into software engineering. The goal is to include the ability to consider abnormal scenarios into software development. More specifically, misuse behavior is described in collaboration diagrams. These are supported by a formal semantics given as positive resp. negative graphical constraints that are based on typed and attributed graphs. It allows one to detect and remove redundancies and conflicts.

**B. Beckert and G. Beuster (University of Koblenz): Formal Specification of Security-Relevant Properties of User Interfaces**

This paper explains how to formally model security relevant properties of user interfaces using the Object Constraint Language (OCL) at the hand of, firstly, the input-output functionality of an operating system. Secondly, using a text-based email system as an example, it is explained how input-output-related security properties of an application can be formally specified with the goal of formal verification.

**E. Beato (Universidad Pontificia de Salamanca); M. Barrio-Solórzano, C. E. Cuesta, and P. de la Fuente (Universidad de Valladolid): Verification Tool for the Active Behavior of UML**

This paper presents a verification approach for behavioral specifications in UML. The model-checker SMV is used to verify properties of systems specified in UML class diagrams, statecharts and activity diagrams.

**M. Bujorianu (University of Cambridge) and M. C. Bujorianu (University of Kent): Towards Engineering Development of Distributed Stochastic Hybrid Systems with UML**

This work investigates the possibilities to provide a bridge between systems control engineering and software engineering using UML. Specifically, it is examined how to use UML to model stochastic hybrid systems.

**J. Knudsen, R. Gottler, M. Jacobsen, M. W. Jensen, J. G. Rye-Andersen, and A. P. Ravn (Aalborg University): Integrating an UML Tool in an Industrial Development Process - A Case Study**

This paper reports on experiences with integrating a UML CASE tool into an industrial software development process in the field of embedded systems. As part of the process, the application documentation, test specifications and program code are kept in the tool so that the tool is fully integrated into the development process.

**M. Sand (University of Erlangen-Nuremberg): Verification and Test of Critical Systems with Patterns and Scenarios in UML**

This work introduces an approach for verification and test of critical hardware near embedded systems. The approach aims to provide automated tools for the simulation or verification of the models against requirements. These requirements, as well as standardized general solutions for providing them, are introduced using patterns and scenarios.

**K. Tabata, K. Araki, S. Kusakabe, and Y. Omori (Kyushu University): A Case Study of Software Development Using Embedded UML and VDM**

The goal of the work sketched in this paper is to design a software development method for embedded systems which combines the formal method VDM++ with Embedded UML, which is a model-based software development method that incorporates best practices and specialized processes from embedded system development.

**O. Tahir, C. Sibertin-Blanc, J. Cardoso (Université Toulouse): A Semantics of UML Sequence Diagrams Based on Causality Between Actions**

The goal of the work sketched in this paper is to propose a semantics for the UML sequence diagrams based on a relation of causality between the actions of emission and reception of messages. A particular interest of the research lies in the relationships between different kinds of behavioral diagrams, such as statecharts and sequence diagrams.

## 3   Conclusion and Future Work

As a conclusion that can be drawn from the success of the workshop, it seems that the topic of critical systems development using modeling languages such as UML is enjoying increasing attention. As was reported in several of the talks, there is an increasing number of industrial projects making increasingly sophisticated use of UML or similar notations for developing systems that have to satisfy intricate critical requirements. This shows that the various obstacles that were mentioned are not unsurmountable, but that, in fact, in various application scenarios people have managed to deal with them. Therefore, it seems that the use of UML and related notations is beginnig to reach a level of maturity which allows serious industrial use.

On the other hand, the various discussions at the workshop, during the panel session, the talks, the coffee breaks, and the workshop dinner also showed that, although the topic of critical systems development using modeling languages such as UML is seen to be a very worthwhile and timely one, many challenges still remain. The various issues that were touched and were it was felt that the final answer has not been found yet include:

- the development of sophisticated tool-support (such as automated theorem provers) for critical systems development using modeling languages such as UML,
- in particular, how to find the optimal trade-off between precision and flexibility in a notation such as UML,
- whether the use of such notations for critical systems development scales up sufficiently in an industrial context,
- and whether UML itself is suitable also for describing critical requirements (as opposed to design) or whether something else is needed.

These points already promise to again give an exciting sequel to the CSDUML workshop series in 2005. More information can be found at [CSD05].

The 2004 workshop proceedings with the contributed papers are available as [JFFR04]. As with previous CSDUML workshops, it is planned to edit a special section of the Journal of Software and Systems Modeling (Springer-Verlag) with selected contributions of the workshop. Up-to-date information on this and the workshop can be found at the workshop home-page [CSD04].

*Acknowledgements.* We would like to express our deepest appreciation especially to the invited speaker Connie Heitmeyer and also to the panelists and to the authors of submitted papers for their valuable contribution. We thank no less the program committee members and the additional referees for their expertise. We would also like to thank the UML'04 conference chair Ana Moreira (New University of Lisbon), the workshop chair Ambrosio Toval (University of Murcia), the local arrangements chair Isabel Sofia Brito (Politécnico de Beja), and the students involved in the organization at TU Munich (in particular Dimitri Kopjev and Britta Liebscher) for their indispensable help. In addition, some of the organizers thank their various projects (including the Verisoft project) for funding.